\documentclass[12pt,fleqn,letterpaper]{article}
\pdfoutput=1
\usepackage{amssymb,amsmath}
\usepackage{graphicx}
 \usepackage{url} 
 \usepackage{hyperref}

\begin{document}
\addcontentsline{toc}{section}{Title and abstract}
\begin{center}

\Large{\bf The emergence of cosmic repulsion   } \\

\small{Gregory Ryskin \\
\it{Robert  R. McCormick School of Engineering and Applied Science, Northwestern University, Evanston, Illinois 60208, United States} \\}
\end{center}

\begin{abstract}

In cosmology based on general relativity, the universe is modeled as a fluid. The transition from the Einstein field equation to its large-scale (cosmological) version is thus analogous to the transition, for a system consisting of a large number of molecules, from the molecular/kinetic description to thermodynamics and hydrodynamics. The cosmic fluid is an effective continuum defined on the cosmological scales (only); for such a continuum, the appearance of new emergent properties should be expected.  (Emergence of space-time and gravity is not discussed here.)  When these new properties are calculated, the following predictions result: (a) the universe is spatially flat; (b) its expansion  is accelerating; (c) dark energy makes up 75\% of the total energy density of the universe; (d) the pressure of dark energy is equal and opposite to its density.  All of these are in good agreement with the observational data. Also in favor of the present model are the absence of adjustable parameters, and consistency with the second law of thermodynamics. The distance-redshift  relation  predicted by the model is in   good agreement with the  Hubble diagram of Type Ia supernovae.

\end{abstract}

\section{Introduction}\label{intro}

In cosmology based on general relativity, the universe is modeled as a fluid. The transition from the Einstein field equation to its large-scale (cosmological) version is thus analogous to the transition, for a system consisting of a large number of molecules, from the molecular/kinetic description to the effective-continuum description of thermodynamics and hydrodynamics. An important consequence of such transitions -- changes of the level of description -- is the appearance of new emergent properties such as temperature or pressure, which have no meaning before the transition.
  
An instrument suitable for measuring macroscopic variables does not ``see'' the molecules, whereas a microscopic instrument (or an imaginary being), capable of measuring the velocities of individual molecules and the forces between them, would not be able to ``feel'' the pressure.  ``A precise determination of temperature is incompatible with a precise determination of the positions and velocities of the molecules"  --  Niels Bohr  \cite{Heisenberg}. Concepts that are well defined on the macroscale become meaningless on the microscale, and vice versa. 

These purely classical  issues were  discussed by Niels Bohr  when he was developing the  concept of complementarity in quantum theory; he used the term ``complementarity" in reference to these classical  issues as well  \cite{Bohr}.  The two types of complementarity are, of course,  entirely distinct.  The present discussion does not involve quantum theory in any way.
 
The cosmic fluid that serves as the model of the universe is an effective continuum defined on cosmological scales (only); for such a continuum, the appearance of new emergent properties should be expected.  (Emergence of space-time and gravity is not discussed in the present work.)  The ``microscale'' in this case includes everything from atoms and molecules to galaxy clusters; thus the emergent properties of the cosmic fluid will not be directly accessible to laboratory experiments or astronomical observations, but will be manifest only through their effect on the dynamics of the universe as a whole.

When these new properties are calculated, the following predictions result: (a) the universe is spatially flat; (b) its expansion  is accelerating; (c) dark energy makes up 75\% of the total energy density of the universe; (d) the pressure of dark energy is equal and opposite to its density.  

All of these  are in good agreement with the observational data. The above summary of the results is phrased in terms of dark energy solely for convenience; in fact, the present model eliminates the dark energy concept altogether -- the emergent properties of the cosmic fluid suffice to explain the dynamics of the universe.

The present model  differs fundamentally from the standard model of cosmology  ($\Lambda$CDM), whatever the choice of the parameters in the latter. (The present model has no adjustable parameters.)  For example, the present model implies that the expansion has been accelerating ever since the end of the radiation-dominated era. In the standard model, such a scenario would preclude the growth of structure that eventually led to the formation of galaxies. Not so in the present model: the growth of structure does  occur, and its rate is in reasonable agreement with the data (Section \ref{growth}).

As mentioned above, the present model is in good agreement with all the major conclusions of observational cosmology.  Detailed comparison with observations will become possible after a complete cosmology is built on the new basis; this is far beyond the scope of  the present work.  One  comparison that can be made now -- with the  Hubble diagram of Type Ia supernovae  -- shows  good agreement (figure \ref{fig1} and Section \ref{flatacceldark}).

Assuming the data continue to support it, the present model  has a number of attractive features. These include the absence of adjustable parameters, and the elimination of the dark energy concept. The model resolves three  problems of cosmology -- the flatness problem, the cosmological constant problem, and the coincidence problem.  Its consistency with the second law of thermodynamics explains the large-scale homogeneity of the universe. It even sheds new light on the origin of inertia.

The analysis (Sections \ref{cosmofe},\ref{mean},\ref{potential}) is based entirely on general relativity, without the cosmological constant. No additional fields are invoked or introduced; quantum aspects are not discussed.  The mean-field approach, borrowed from statistical physics, leads to the desired result with a minimum of calculation. The predictions of the model are compared with observational data in Sections \ref{flatacceldark},\ref{growth}; consistency with the second law of thermodynamics is discussed in Section \ref{thermo}. A closely related statistical-physics problem is reviewed in the Appendix.

\section{The cosmological field equation}\label{cosmofe}
I follow the notational conventions of \cite{MTW}; in particular, both the speed of light $c$ and the Newtonian gravitational constant $G$ are set to 1. Thus time, mass, and energy are measured in units of length; energy density and pressure have dimensions of (length)$^{-2}$.

	The Einstein field equation of general relativity takes the form 
\begin{equation}\label{efe}{{R}_{\mu \nu }}-\frac{1}{2}{{g}_{\mu \nu }}R=8\pi {{T}_{\mu \nu }}.\end{equation}
Here Greek indices run over 0,1,2,3; the coordinate ${{x}^{0}}$ is time $t$; the spatial coordinates are ${{x}^{1}},{{x}^{2}},{{x}^{3}}$;  ${{g}_{\mu \nu }}$ is the metric tensor, which determines the geometry of space-time and plays the role of gravitational field; ${{R}_{\mu \nu }}$ is the Ricci curvature tensor constructed from ${{g}_{\mu \nu }}$;  $R={{R}_{\alpha \beta }}{{g}^{\alpha \beta }}$ is the Ricci   scalar; and ${{T}_{\mu \nu }}$ is the stress-energy tensor of matter, including radiation and non-gravitational fields. Summation over repeated indices is implied. The metric signature is $-+++$ .

	If the complete solution were known of the Einstein equation (\ref{efe}) everywhere in the universe, all the gravitational phenomena, from the fall of an apple to black holes to the expansion of the universe, would be precisely described. Such a solution will never be known. Beginning with Einstein's paper of 1917  \cite{Einstein 1917}, the cosmological implications of general relativity are deduced from a different (though identical in form) equation, viz.,		\begin{equation}\label{cosmee}{\mathsf{R}_{\mu \nu }}-\frac{1}{2}{\mathsf{g}_{\mu \nu }}\mathsf{R}=8\pi {\mathsf{T}_{\mu \nu }},\end{equation}
where ${\mathsf{g}_{\mu \nu }}$, ${\mathsf{R}_{\mu \nu }}$, and $\mathsf{R}$  describe the \emph {large-scale}  geometry of the space-time; ${\mathsf{T}_{\mu \nu }}$  is the stress-energy tensor of the homogenized (smoothed-out) matter distribution.

	Equation (\ref{cosmee}) cannot be derived by averaging equation (\ref{efe}) because the left-hand side of equation (\ref{efe}) is nonlinear.  (Physically, the nonlinearity means that gravitational field itself acts as a source of gravity.)  Equation (\ref{cosmee}) is thus an additional hypothesis, the cosmological field equation, modeled on the Einstein field equation (\ref{efe}). This fact is well known, and continues to be a cause of a considerable controversy; see the recent reviews \cite{Clarkson, Ellis 2011}  and references therein.  I shall not review this literature here because my approach differs fundamentally from the previous work.  The following excerpt \cite[p.~4]{Clarkson} summarizes the issue: 

``Now the Einstein equations may be assumed to hold at the `local' scale: after all, this is the scale where they have been exquisitely checked \ldots But the averaging process  \dots does not commute with evaluating the inverse metric, connection coefficients, Ricci tensor and Ricci scalar  \ldots Hence if the [Einstein field equations] hold at `local' scale   \ldots they will not hold at [cosmological] scales  \ldots  In essence, it is an \emph{assumption} that Einstein's equations also hold for an averaged geometry, as well as a local one."

	The physical system whose dynamics the cosmological field equation purports to describe is the cosmic fluid -- an effective-continuum representation of the universe together with its matter content.  Experimental and observational tests of the Einstein field equation have, strictly speaking, no bearing on the validity of equation (\ref{cosmee}); only the observational cosmology has. The fact that the Einstein field equation is in excellent agreement with observations does suggest that something like equation (\ref{cosmee}) will be the correct mathematical description of cosmology. But that description has to be derived in a rational manner, not simply taken over from a different physical situation described by equation (\ref{efe}).

	As already mentioned, no averaging procedure can be devised that would lead from equation (\ref{efe}) to equation (\ref{cosmee}). Moreover, it is highly unlikely that any kind of averaging of the Einstein field equation would lead to the correct cosmological field equation. (At this point the present approach begins to diverge from the previous work, reviewed in \cite{Clarkson, Ellis 2011}.)  Generally speaking, averaging is not the right approach when the appearance of emergent properties is expected. The case in point is thermodynamics, where emergent properties of an effective continuum, such as temperature or pressure, cannot be obtained by averaging -- the microscale analogs of these properties do not exist. The greatest challenge present the emergent properties, such as entropy or negative pressure, that are not directly observable (measurable). Thermodynamics resolves this conundrum in a radical manner: the microscale is deliberately ignored, new principles are set forth that involve the emergent properties, and a quantitative theory is built; the whole construction is validated by comparison with experiment.

	Nothing so radical will be required in order to obtain the correct cosmological field equation.  The fundamental principles of general relativity -- as opposed to their particular realization in the Einstein field equation -- are sufficiently wide-ranging to apply to the cosmic fluid as well. Thus the route to the cosmological field equation is simple: 

	Instead of following the form of the Einstein field equation, follow the logic of general relativity that led to its formulation.

	In particular, respect ``the requirement that the energy of the gravitational field shall act gravitatively in the same way as any other kind of energy'', a major factor in the original formulation \cite[\S16]{Einstein 1916}.

	The cosmological-scale metric ${\mathsf{g}_{\mu \nu }}$  does not describe gravitational fields on any smaller scales, be it laboratory or galaxy cluster. Consider a homogeneous and isotropic universe which is spatially flat; the curvature of the space-time defined by ${\mathsf{g}_{\mu \nu }}$ is then completely determined by the scale factor of the universe $a(t)$ and its first and second time derivatives.  Thus the gravitational fields on all scales smaller than the cosmological one are ``invisible'' in the geometric, left-hand side of equation (\ref{cosmee}).  

	Does this invisibility imply that the energy of these gravitational fields is missing from the description provided by equation (\ref{cosmee})?  (The elastic energy of  a piano string, say, includes contributions of all harmonics, not only of the first.)  If so, the missing gravitational energy should be evaluated, and the corresponding stress-energy tensor added to that of the other sources on the right-hand side of equation (\ref{cosmee}).

	At the first sight, the answer to the above question is clearly in the affirmative. The gravitational binding energy -- the energy of the gravitational interactions between the constituent parts of a matter distribution -- can be estimated as the total mass of the distribution $ M$   times the Newtonian gravitational potential  ${{\phi }_{N}}=O(-{M}/{L})$, where $L$  is the characteristic size of the distribution. Homogenization of that matter distribution over a large volume (of dimensions much greater than $L$) conserves the total mass $M$, but essentially eliminates the gravitational binding energy; thus the latter should be added to the right-hand side of equation (\ref{cosmee}).

	This is incorrect. And the reason is not the smallness of the gravitational potential.

	The reason is that the astronomically observed mass-energy $M$ of a matter distribution, the source (the ``input data'') for the stress-energy tensor ${\mathsf{T}_{\mu \nu }}$  in equation (\ref{cosmee}), already includes the energy of the gravitational interactions between the constituent parts of the distribution. This is so because the astronomical observations that are used to infer $M$, depend on the \emph{gravitational} effect of the matter distribution, and involve distances much greater than $L$. That is, they measure, from an asymptotically flat (or nearly so) space-time, the true gravitating mass-energy. Examples are measurements based on Kepler's laws, and on gravitational lensing.  If other types of observations lead to a different value for $M$, that value is not trusted -- recall the discovery of dark matter.

	There is thus no need to include ``by hand'' on the right-hand side of equation (\ref{cosmee}) the energy of gravitational interactions within a star, within a galaxy, within a galaxy cluster, \dots  

	It is tempting to finish the last sentence with ``and so on \emph{ad infinitum}''.  But this would be wrong:  the above logic fails when the level of the entire universe is reached.   

For the gravitational interaction between a local matter distribution and the rest of the universe, there exist no distances greater than $L$, and no asymptotically flat space-time from which to measure mass-energy \cite[\S19.4]{MTW}.  (The range of quasi-static gravitational forces is not limited by causality.  These forces do not transmit information; they act through the constraint  equations of general relativity, which are instantaneous equations satisfied on each  space-like hypersurface  \cite{Ellis_Sciama}.)  Therefore astronomical observations cannot sense the energy of this interaction, except indirectly, through its effect on cosmology.  This interaction (called ``cosmic interaction'' hereafter) is completely dominated by the distant matter, so the local density variations (inhomogeneities) are irrelevant.  And the energy of the cosmic interaction can be very large, as the following consideration suggests.

	To find the energy of gravitational interaction between a test mass and the rest of the universe, one needs the Newtonian gravitational  potential due to the latter -- but this potential is infinite.  This means that the Newtonian approximation fails when the level of the entire universe is reached. This also suggests that in general relativity the effect  will be very large (fully relativistic).  

	Thus the energy of the gravitational interaction between a local matter distribution and the rest of the universe is missing, for fundamental reasons, from the mass-energy data provided by astronomical observations and represented by ${\mathsf{T}_{\mu \nu }}$.  To complete the formulation of the cosmological field equation, the stress-energy tensor of the cosmic interaction, denoted  ${{\Phi }_{\mu \nu }}$, must be added to ${\mathsf{T}_{\mu \nu }}$.  The correct cosmological field equation is 
\begin{equation}\label{newcosme}{\mathsf{R}_{\mu \nu }}-\frac{1}{2}{\mathsf{g}_{\mu \nu }}\mathsf{R}=8\pi{( {\mathsf{T}_{\mu \nu }}+{{\Phi }_{\mu \nu }})}.\end{equation}
The sum ${ {\mathsf{T}_{\mu \nu }}+{{\Phi }_{\mu \nu }}}$ is the stress-energy tensor of the cosmic fluid.  Sections \ref{mean} and \ref{potential} are devoted to the calculation of ${{\Phi }_{\mu \nu }}$, using   the mean-field approach borrowed from statistical physics. 

It is important to emphasize that the present work is based entirely on general relativity. No additional ingredients (new fields,  new types of interactions, etc.) are introduced.  In contrast to equation (\ref{cosmee}), the cosmological field equation (\ref{newcosme}) differs in form from the Einstein field equation (\ref{efe}) --  because   application of general relativity on cosmological scales  requires  a different mathematical representation of the same physical content.  Note that   tensor ${{\Phi }_{\mu \nu }}$  in equation (\ref{newcosme}) will be  determined, via a self-consistent limiting procedure of the mean-field theory, entirely on the basis of the Einstein field equation (\ref{efe}).  Thus the final version of  equation (\ref{newcosme}), the version that will be  used in cosmological calculations,  is  implied, fully and  rigorously, by the Einstein field equation.

An analogy can be drawn between the present problem and the problem of  turbulence   \cite{Tennekes}, where the mean-flow  equations, obtained by averaging the  Navier-Stokes equations,  differ in form from the latter: the mean-flow  equations  contain an extra  ``Reynolds-stress"  term.  (Because the  Navier-Stokes equations are nonlinear.)   The crucial difference is that in the present problem, the large-scale equations can be obtained in closed form, whereas the  closure problem of turbulence  remains unsolved:  The Reynolds stress tensor can be formally expressed via velocity fluctuations (as   $-\rho< u^\prime_i  u^\prime_j>$), but cannot be related to the mean-flow velocity. To obtain a similar formal expression for tensor ${{\Phi }_{\mu \nu }}$ would be extremely difficult, and the example of turbulence shows that this route to the  large-scale equations is unlikely to succeed.  A very different  approach is pursued in the present work.  

The ultimate reasons for the  solvability of the present problem are the universality and the long    range  of  gravitational interaction; these features  guarantee that the mean-field theory works and yields exact results.  By contrast, hydrodynamic interactions in turbulent flow,  mediated by vortex stretching, are  non-universal  and  short range; no mean-field theory is possible.

\section{Mean-field theory for the cosmic fluid}\label{mean}
Consider a homogeneous and isotropic universe in which the energy density of relativistic matter and radiation is negligibly small compared to that of the non-relativistic matter (``dust'').  The kinetic pressure of the non-relativistic matter is much smaller than its energy density ${{\rho }_{m}}(t)$; their ratio is $\sim{{({v}/{c}\;)}^{2}}$, where $v$ is the mean velocity of the dust particles.    Thus the kinetic pressure of matter can be safely neglected.  (The subscript $m$ stands for ``matter'', which will include dark matter; ${{\rho }}(t)$ without a subscript will denote the total energy density of the cosmic fluid, and $p(t)$, its total pressure.)

	But the pressure due to the cosmic interaction may be significant. To find the stress-energy tensor of the cosmic interaction ${{\Phi }_{\mu \nu }}$, its energy density  must be included in the cosmological  Lagrangian.

	To describe the cosmic interaction, I introduce the mean-field potential ${{\phi }_{c}}$, felt by a test mass as a result of gravitational interaction between this test mass and the rest of the universe.  (The subscript $c$ stands for ``cosmic''.)  The mean-field energy density of the interaction is then ${{\rho }_{m}}{{\phi }_{c}}$.  In co-moving coordinates, ${{\phi }_{c}}$ could only depend on time; we shall see in Section \ref{potential} that ${{\phi }_{c}}$ is actually a constant. The value of ${{\phi }_{c}}$ should be negative because gravitational interactions are attractive. Similarly to ${{\rho }_{m}}$, ${{\phi }_{c}}$  is a scalar -- its value does not change under coordinate transformations. This does not violate the principle of equivalence because ${{\phi }_{c}}$ is spatially uniform and locally unobservable -- it is an effective cosmic potential that manifests itself on cosmological scales only.

	Note the analogy with the van der Waals fluid (see Appendix), where the mean field describes the intermolecular interaction between the test particle and all other particles. The pressure due to this attractive interaction is negative (the interaction energy increases when the system expands), and equal to the mean-field energy density of the interaction. Its magnitude can reach thousands of atmospheres (in liquid state).

	Returning to cosmology, let ${{p}_{\phi}}$  denote the pressure due to the cosmic interaction.  By analogy with the van der Waals fluid, ${{p}_{\phi}}$  is expected to be negative, equal to ${{\rho }_{m}}{{\phi }_{c}}$.  It will be seen below that ${{p}_{\phi}}$  is very large, comparable in magnitude to ${{\rho }_{m}}$. This reflects the fact that the cosmic interaction involves the universe as a whole.

	The mean-field theory neglects local density fluctuations, and fails when such fluctuations become large and correlate over distances comparable to the range of interactions (as happens, e.g., near the critical point of a phase transition).  Conversely, the mean-field theory becomes exact for  interactions of very long range: The longer the range of interaction, the greater the relative contribution of the distant matter; this contribution is insensitive to local density fluctuations. This fact is not of much use in the theory of phase transitions because the range of intermolecular forces is limited. But it is important for the present analysis.

There is, of course, no hope to calculate ${{\phi }_{c}}$ by integrating the  gravitational interaction energy between a test mass and all other masses in the universe: the interaction energy depends on the distance as ${{r}^{-1}}$, and the integral diverges. 
This is, actually,  a good news: it means that ${{\phi }_{c}}$ is completely dominated by the influence of the far-away masses, so the results of the mean-field theory will be exact. It also means that the Newtonian approximation fails  -- even though  it describes very well the gravitational interaction between any  two  distant objects of finite size.

In general relativity, the divergence will ``saturate'' at a finite value of ${{\phi }_{c}}$.  The nonlinearity of general relativity is crucial here, just as it is in preventing the divergence of gravitational energy for a matter distribution whose size tends to zero -- a black hole is formed instead.

	General relativity will be used in Section \ref{potential} to deduce the value of ${{\phi }_{c}}$, thus achieving the self-consistent closure, within the mean-field theory, of the present model. An estimate of ${{\phi }_{c}}$ by an order of magnitude would be useful at this point; for a hypothetical finite (spatially closed) universe it can be stated as follows
\begin{equation}\label{estimate_phi}{{\phi }_{c}}\sim-(\text{mass of the universe)/(radius of the universe)}.\end{equation}
For a finite universe, the ratio of its mass to its radius is thought to be $O(1)$ \cite{Einstein 1917}, \cite[p.~548]{MTW}.  A rough estimate for ${{\phi }_{c}}$ is then $O(-1)$, an extremely large (fully relativistic) value. For comparison, the gravitational potential at the surface of the Sun is $O(-{10}^{-6})$,  of a white dwarf, $O(-{10}^{-4})$,  of a neutron star, $O(-{10}^{-1})$.  It becomes  $O(-1)$  only in the vicinity of a black hole.

	Since ${{\phi }_{c}}$ is uniform throughout the universe, its magnitude is not manifest; the only observable effect of ${{\phi }_{c}}$ is the cosmological one. In fact, both ${{\phi }_{c}}$ and the associated pressure ${{p}_{\phi}}$  are the cosmological scale concepts that become meaningless on the scales accessible to astronomical observations or physical experiments, just as in ordinary fluids the concepts of temperature or pressure become meaningless on the molecular scales.  (May it be possible, nevertheless, to devise an experiment that would detect and measure  ${{\phi }_{c}}$?)

	To derive an expression for ${{\Phi }_{\mu \nu }}$, equation (\ref{newcosme}) should be viewed as arising from the  least action principle, with the interaction energy density ${{\rho }_{m}}{{\phi }_{c}}$ included in the cosmological Lagrangian.  
In general relativity, the Lagrangian  is  a sum  of two parts,
\begin{equation}\label{L}{\mathcal L = \mathcal{L}_{geom} + \mathcal{L}_{field}},\end{equation}
where $\mathcal{L}_{geom}$ is the Ricci scalar (times a constant), while $\mathcal{L}_{field}$ includes contributions for  all "matter fields" -- more precisely, for all fields not included in  $\mathcal{L}_{geom}$   \cite[\S\S21.1-21.3]{MTW}.  The stress-energy tensor of the  fields   in $\mathcal{L}_{field}$  can be calculated as \cite[\S21.3]{MTW}
\begin{equation}\label{TfromL}{-2\frac{\delta \mathcal{L}_{field}}{\delta \mathsf{g}^{\mu \nu }} + \mathsf{g}_{\mu \nu }\mathcal{L}_{field}}.\end{equation}
The mean-field  energy density ${{\rho }_{m}}{{\phi }_{c}}$ of the cosmic interaction  should now be included in the cosmological Lagrangian. (In which  $\mathcal{L}_{geom}$ is  the Ricci scalar ${\mathsf{R}}$ of the large-scale geometry.) This means the appearance in $\mathcal{L}_{field}$ of an additional term,
\begin{equation}\label{Lphi}{\mathcal{L}_{\phi} = {{\rho }_{m}}{{\phi }_{c}}}.\end{equation}
Since $\mathcal{L}_{\phi}$ is independent of $\mathsf{g}^{\mu \nu }$, it follows from (\ref{TfromL}) that the stress-energy tensor of the cosmic interaction is  $\mathcal{L}_{\phi}$ times the metric tensor of the large-scale geometry, i.e.,
\begin{equation}\label{Phi}{{\Phi}_{\mu \nu }}={{\rho }_{m}}{{\phi }_{c}}{\mathsf{g}_{\mu \nu }}.\end{equation} 
The pressure due to ${{\phi }_{c}}$ is then\begin{equation}\label{}{{p}_{\phi }}={{\rho }_{m}}{{\phi }_{c}},\end{equation} 
as expected on physical grounds.

	The  stress-energy tensor ${\mathsf{T}_{\mu \nu }}$  of the non-relativistic matter is well known; in co-moving coordinates, only one component is non-zero,  ${\mathsf{T}_{00}}={{\rho }_{m}}$.  The energy density  and the pressure of the cosmic fluid can now be written as\begin{equation}\label{rho}\rho ={{\rho }_{m}}(1-{{\phi }_{c}}),\end{equation}\begin{equation}\label{p}p={{\rho }_{m}}{{\phi }_{c}}.\end{equation} 
The cosmic fluid is an effective-continuum representation of the universe together with its matter content, appropriate on cosmological scales. 

\section{The value of the cosmic potential ${{\phi }_{c}}$}\label{potential}
The ultimate source of information concerning ${{\phi }_{c}}$ is the Einstein field equation (\ref{efe}). It is a tensor equation, while ${{\phi }_{c}}$ is a scalar. This suggests that a scalar equation derivable from the Einstein field equation should be sufficient for deducing the value of ${{\phi }_{c}}$. 

	The only covariant scalar equation that is an exact consequence of the Einstein field equation is the trace of the latter\begin{equation}\label{R=-8piT}R=- 8\pi T,\end{equation} 
where $T={{g}^{\alpha \beta }}{{T}_{\alpha \beta }}$ is the trace of the stress-energy tensor.  Generally, this equation embodies only a small part of the geometric (gravitational) information contained in general relativity; the complete Einstein field equation (\ref{efe}) is equivalent to ten differential equations for the components of the metric (constrained by four Bianchi identities).  However, a homogeneous and isotropic universe possesses a high degree of symmetry; the Einstein field equation reduces to two ordinary differential equations (the Friedmann equations).  Furthermore, the energy-conservation equation, which can be derived from the Friedmann equations, is actually independent: it is the first law of thermodynamics.  (The Einstein equation is so constructed that this law is automatically satisfied, by virtue of Bianchi identities.)  Combining the energy-conservation equation with the first Friedmann equation, one can derive the second Friedmann equation \cite[\S27.8]{MTW}. Thus, in the Friedmann universe, the entire geometric (gravitational) content of general relativity reduces to a single ordinary differential equation.  The same content should then be fully conveyed by the scalar equation (\ref{R=-8piT}).

	In this section, I will consider a homogeneous and isotropic Friedmann universe that is spatially flat (Euclidean).  This case is the simplest one; also, the observations suggest that the universe is, in fact, spatially flat.  The other two possibilities will be discussed in the next section. 

	Using co-moving coordinates, the line element in the spatially flat case can be written as \begin{equation}\label{line_elem}d{{s}^{2}}=-d{{t}^{2}}+{{a}^{2}}(t){{\delta }_{ij}}d{{x}^{i}}d{{x}^{j}},\end{equation} 
where ${{\delta }_{ij}}$ is the Kronecker delta; Latin indices run over 1,2,3; ${{x}^{j}}$ are dimensionless Cartesian coordinates; and $a(t)$ is the scale factor, with dimensions of length. The scale factor's dependence on time -- the expansion of the universe -- determines the curvature of the space-time.

	If the time coordinate is redefined so that $dt=a(t)d{t}'$, the line element (\ref{line_elem}) becomes\begin{equation}\label{conform}d{{s}^{2}}={{a}^{2}}(-d{{{t}'}^{2}}+{{\delta }_{ij}}d{{x}^{i}}d{{x}^{j}}).\end{equation} 
This shows that the space-time is conformally flat, $a({t}')$  being the conformal factor.

In a conformally flat space-time, equation (\ref{R=-8piT}) can be reformulated, rigorously, as a field equation for a scalar gravitational field  $\phi$ (identified with the conformal factor) on a flat space-time with metric ${{\eta }_{\mu \nu }}=\text{diag}(-1,1,1,1)$.  The result is (\cite[Exercise 17.8]{MTW},\cite[p.~42]{Hawking Ellis}) \begin{equation}\label{dal}\Box\phi=\frac{4\pi }{3}\phi {{T}_{flat}},\end{equation} 
where $\Box$ is the flat-space-time d'Alembertian operator, $ \Box\phi \equiv {{\eta }^{\alpha \beta }}\phi {{,}_{\alpha\beta }}$, and ${{T}_{flat}}={{\eta }^{\alpha \beta }}{{T}_{\alpha \beta }}$. Equation (\ref{dal}) is the exact consequence of the Einstein field equation (\ref{efe}) combined with the symmetry of the Friedmann universe.  (With ${{g}_{\mu \nu }}={{a}^{2}}{{\eta }_{\mu \nu }}$ as in equation (\ref{conform}), and  $\phi =a({t}')$,  equation (\ref{dal}) reduces to an ordinary differential equation -- the sum of the two Friedmann equations.)  That is, in spite of the fact that a scalar field is being used to describe gravity, equation (\ref{dal}) expresses the same content as is implied by general relativity, under the restrictions imposed by the symmetry of the space-time.

	In this section the purpose is to deduce, by self-consistency, the value of the cosmic potential ${{\phi }_{c}}$ introduced in Section \ref{mean}. Since the result will be used to infer the dynamics of the universe as a whole, the self-consistency condition must arise from local gravitational physics (to avoid a circular argument). With this goal in mind, consider some local distribution of matter with non-relativistic density ${{\rho }_{m}}({{x}^{\alpha }})$, moving with non-relativistic velocities; then $T=-{{\rho }_{m}}$  and  ${{T}_{flat}}=-{{\rho }_{m}}{{a}^{2}}$.   If equation (\ref{dal}) were applicable, the time derivative in the d'Alembertian would be negligibly small compared to the spatial derivatives, and equation (\ref{dal}) would become, to a very good approximation,\begin{equation}\label{lapl}\nabla^{2}\phi =-\frac{4\pi }{3}\phi {{\rho }_{m}},\end{equation} 
where derivatives in the Laplacian are taken with respect to the dimensional Cartesian coordinates $a{{x}^{j}}$.  (In the d'Alembertian of equation (\ref{dal}), derivatives are taken with respect to the dimensionless coordinates  ${{x}^{\alpha }}$.)  Equation (\ref{lapl}) does not change if  $\phi$ is multiplied or divided by an arbitrary function of time; thus, in what follows, $\phi$ in equation (\ref{lapl}) will be viewed as a dimensionless function of position.

	However, the dependence of  ${{\rho }_{m}}({{x}^{\alpha }})$ on position, and the variation of $\phi$ caused by it,  contradict the symmetry assumptions that led to equation (\ref{dal}). Hence equation (\ref{lapl}) is an extension of equation (\ref{dal}) beyond its limits of validity; is such an extension permissible?

	Generally speaking, it is not. As is well known, gravity cannot be described by a scalar field on a flat space-time; in particular, the deflection of light cannot be described \cite[p.~71]{Hawking Ellis}.  Massless particles such as photons cannot couple to a scalar gravitational field -- they can neither feel it, nor be a source of it (note that  $T = 0$ for radiation).

	As soon as the symmetry that led to equation (\ref{dal}) is not complete, only some aspects of gravity may be approximately described by a scalar field.  Equation (\ref{dal}) is then only approximate,  becoming an exact consequence of the Einstein field equation in the limit in which the symmetry is restored. 

	Our goal, however, is to deduce a general property of $\phi$, viz., the value of  ${{\phi }_{c}}$, to be used in the cosmological-scale calculations for the Friedmann universe; such calculations are, of course, in compliance with the required symmetry.  It should then be possible to use equation (\ref{dal}) as an approximate description of gravity if the space-time is sufficiently close to the required symmetry; with an implication that the limit is then taken in which the symmetry is restored.  (The variation of $\phi$   due to  ${{\rho }_{m}}({{x}^{\alpha }})$ is to be viewed as virtual and infinitesimal.) The  result of such an approximate treatment should be the same as what would be found if the exact  Einstein field equation were used before taking the limit, because (a) the equations \emph{per se} are equivalent in this limit, and (b) the limit is not a singular one -- the equations are  of the same  differential order.

A mechanical analogy may be useful at this point.  Consider a thin piano string under high tension, in static equilibrium. The shape of the string is known (the straight line), but this fact is not sufficient to determine the magnitude of the tension. To find the tension, the shape must be perturbed slightly, e.g., by attaching to the string (held horizontally) a small weight at the midpoint. The midpoint displacement divided by the string length is then one-fourth of the ratio of the weight to the tension, so the tension is found easily. Note that this analysis neglects the bending stiffness of the string. The bending stiffness is irrelevant as long as the string remains straight; neglecting it for a deformed string amounts to extending the description appropriate for the high-symmetry case (the straight string) beyond its limits of validity. For high tension and small displacements, such an extension is permissible, and the magnitude of the tension can be determined.

	With equation (\ref{lapl}) justified by the above considerations, recall the experimental fact that the gravitational field of the non-relativistic matter distribution ${{\rho }_{m}}({{x}^{\alpha }})$ is also well described by the Newtonian theory\begin{equation}\label{newt}{{\nabla^{2}\phi }_{N}}=4\pi {{\rho }_{m}},\end{equation} 
where $|{{\phi }_{N}}|<<1$.  Equations (\ref{lapl}) and (\ref{newt}) can be compatible only if \begin{equation}\label{compat}\phi =-3+{{\phi }_{N}}.\end{equation} 
This is the required self-consistency condition.  With  ${\phi }_{N}$  being virtual and infinitesimal, this condition  implies the following value for the cosmic potential ${{\phi }_{c}}$ introduced in Section \ref{mean}
\begin{equation}\label{-3}{{\phi }_{c}}=-3.\end{equation} 
The number 3 in equation (\ref{-3}) is an integer -- the number of spatial dimensions of the space-time \cite[p.~42]{Hawking Ellis}.

	The new compatibility requirement (\ref{compat}) is independent of the usual one, which says that in the weak-field limit general relativity must reduce to the Newtonian theory of gravity. (That requirement  determines the coefficient $8\pi$ in the Einstein equation.)  The new  requirement guarantees that the Newtonian theory of gravity remains a good approximation, for non-relativistic densities and velocities, in spite of the presence in the universe of the cosmic potential ${{\phi }_{c}}$ that defines the gravitational interaction of a given mass with the rest of the universe. The new requirement results from combining general relativity with the symmetry of the Friedmann universe and the experimental facts concerning non-relativistic gravity. Thus, the new compatibility is between non-relativistic gravity and general relativity \emph{in the cosmological limit}.  Nonlinearity of general relativity makes it possible for two different compatibility requirements, corresponding to two different limits, to coexist.

Note that tensor ${{\Phi }_{\mu \nu }}$  in  the   cosmological field equation  (\ref{newcosme}) is completely determined once the value of the cosmic potential ${{\phi }_{c}}$ is known (see equation (\ref{Phi})).   As the present section shows, both the existence of the cosmic potential   ${{\phi }_{c}}$, and the precise  value of it,  are implied by the Einstein field equation (\ref{efe}).  In this way,  the   final version of  the cosmological field equation (\ref{newcosme}), the version that will be used in the cosmological calculations below,  is fully determined by the Einstein field equation.

\section{Space-times with spatial curvature}\label{curvature}
The Friedmann space-times with positive or negative spatial curvature are conformally flat as well, though the corresponding conformal transformations are more complicated, and the conformal factor, besides being a function of time, is also a function of spatial coordinates.  Most of the logic of the previous section still applies, but because of the spatial variation of the conformal factor, the new compatibility requirement cannot be satisfied.

I conclude that general relativity, combined with the symmetry of the Friedmann universe and the experimental facts concerning non-relativistic gravity, requires the universe to be spatially flat. 

\section{Implications for cosmology I. Flatness, acceleration, dark energy}\label{flatacceldark}
Observational data suggest that the universe is spatially flat, its content dominated by dark energy which makes up $\sim{73\%}$ of the total. The rest is the non-relativistic matter (including dark matter); the energy density of relativistic matter and radiation is negligible, except in the early universe. With dark energy is associated negative pressure, equal and opposite to its density; this negative pressure acts as a source of anti-gravity, with the result that the expansion of the universe is accelerating. The nature of dark energy remains unexplained.

	The amount of astronomical data that have a bearing on cosmology is very large and growing. It is beyond the scope of this paper to build a complete cosmology on the basis of the present model, and to compare its predictions with the available data. The assessment presented below (this section and Section \ref{growth}) will be, of necessity, incomplete.

 Let us begin by testing the present model vs. the major conclusions of observational cosmology, distilled from the multiple types of data: the flatness of the universe, the fact that its expansion is accelerating, the fraction of the total energy density that is attributed to dark energy, and the equation of state of the latter.

	The conclusion of Section \ref{curvature} that the universe must be spatially flat on cosmological scales is in agreement with the observational data.  

	The key result of the present work, equation (\ref{-3}), leads to the following properties of the cosmic fluid (see equations (\ref{rho},\ref{p})),\begin{equation}\label{rho=4}\rho =4{{\rho }_{m}},\end{equation} 
	\begin{equation}\label{p=-3}p=-3{{\rho }_{m}}.\end{equation} 

The part of the stress-energy of the cosmic fluid represented by the interaction tensor ${{\Phi }_{\mu \nu }}$  thus contributes 3/4 of the total energy density.  The pressure of that part is equal and opposite to its energy density.  Both predictions are in good agreement with the observational data, with ${{\Phi }_{\mu \nu }}$  replacing dark energy. 

	The cosmic fluid (overall) equation-of-state parameter $w$, defined by $p=w\rho $, is \begin{equation}\label{-3/4}w=-\frac{3}{4}.\end{equation} 
The equation-of-state parameter $w$ is defined here for the cosmic fluid as a whole. It is not necessary to introduce a separate equation-of-state  parameter for dark energy: dark energy is absent in the present model, being replaced by ${{\Phi }_{\mu \nu }}$.  (If a separate equation-of-state parameter  were  defined  for   ${{\Phi }_{\mu \nu }}$ alone, its value would be  $-1$.)

Note that a   different  notational convention is  prevalent in the  literature:  the symbol $w$ without a subscript is   used  as a generic notation for an equation-of-state parameter of each component separately, e.g., $w=0$ for non-relativistic matter,  $w=1/3$ for  relativistic matter and radiation, $w=-1$ for cosmological constant.  In the recent literature on  cosmology, $w$ usually stands for $w_{de}$,  the    equation-of-state parameter  for   dark energy alone.  

Equations (\ref{rho=4},\ref{p=-3},\ref{-3/4}) apply when the energy density of radiation and relativistic matter is negligibly small compared to the energy density ${{\rho }_{m}}$  of the non-relativistic matter. The kinetic pressure of matter has been neglected.  If it were included, the numerical factor in equation (\ref{p=-3}) would be greater than $-3$, but  less than $-8/3$.  Thus $w$ satisfies the inequality\begin{equation}\label{-3/4w-2/3}-\frac{3}{4}\le w<-\frac{2}{3}.\end{equation} 

It remains to see whether the model predicts accelerating expansion. In a homogeneous and isotropic, spatially-flat universe, the cosmological field equation reduces to the two Friedmann equations \cite{Friedman}, governing the evolution of the scale factor of the universe $a(t)$,\begin{equation}\label{fried1}{{\left( \frac{{\dot{a}}}{a} \right)}^{2}}=\frac{8\pi }{3}\rho ,\end{equation} 
	\begin{equation}\label{fried2}\frac{{\ddot{a}}}{a}=-\frac{4\pi }{3}(\rho +3p).\end{equation} 
The first law of thermodynamics takes the form\begin{equation}\label{firstlaw}d(\rho {{a}^{3}})=-pd({{a}^{3}}).\end{equation} 
Given equations (\ref{rho=4},\ref{p=-3}), the Friedmann equations become\begin{equation}\label{fr1new}{{\left( \frac{{\dot{a}}}{a} \right)}^{2}}=\frac{32\pi }{3}{{\rho }_{m}},\end{equation} 
	\begin{equation}\label{fr2new}\frac{{\ddot{a}}}{a}=\frac{20\pi }{3}{{\rho }_{m}}.\end{equation} 
The first law of thermodynamics becomes\begin{equation}\label{firstlaw_new}d({{\rho }_{m}}{{a}^{3}})=\frac{3}{4}{{\rho }_{m}}d({{a}^{3}}),\end{equation} 
and implies a weak dependence of the energy density of matter on the scale factor,\begin{equation}\label{rho-3/4}{{\rho }_{m}}\propto {{a}^{-{3}/{4}\;}}.\end{equation} 
The Friedmann equation (\ref{fr1new}) then predicts accelerating expansion of the universe,
\begin{equation}\label{a(t)}a(t)\propto {{t}^{{8}/{3}\;}}.\end{equation} 
The Hubble parameter  $H(t)\equiv {{\dot{a}(t)}}/{a(t)}\;$  is  the relative rate of expansion of the universe at time $t$.   Equation (\ref{a(t)}) implies
\begin{equation}\label{3/8}{H(z)}={{H}_{0}}{{(1+z)}^{{3}/{8}\;}},\end{equation}
where $z$ is the cosmological redshift, $1+z={{{a}_{0}}}/{a(t)}$, and the subscript 0 indicates the present time, $z = 0$.  The present-time value ${{H}_{0}}$ of the Hubble parameter is the Hubble constant.

\begin{figure}[h]
\section*{Figures}
\addcontentsline{toc}{section}{Figures}

\centering
\includegraphics[scale=0.8, angle=0]{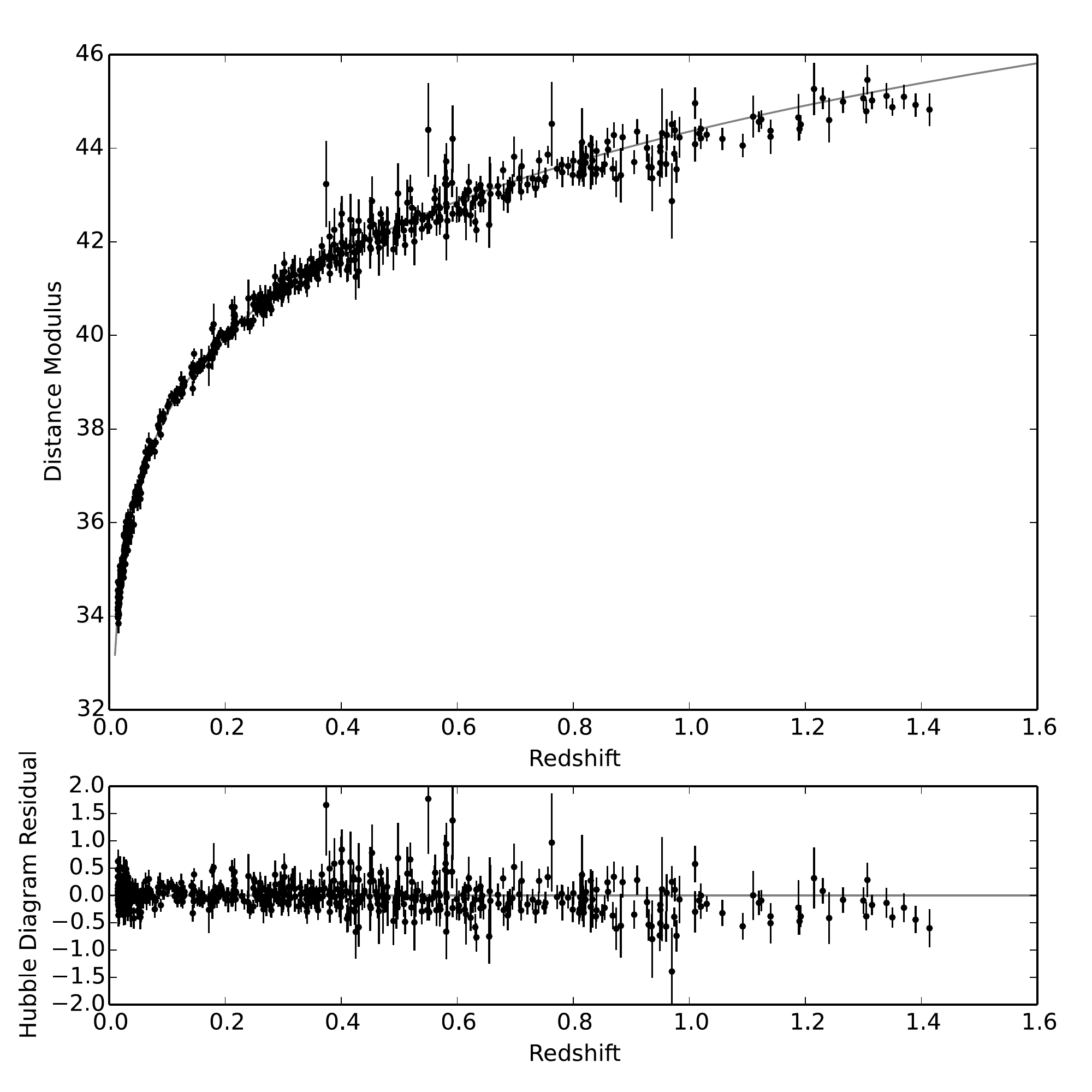}
\caption{The Hubble diagram (distance  vs. redshift) for the Union2.1 compilation of 580 Type Ia supernovae \cite{Suzuki}.  The solid line is the prediction of the present model, equation (\ref{3/8}).}

\label{fig1}
\end{figure}

In figure \ref{fig1} this  prediction  is compared with the observational  data from  Type Ia supernovae, compiled in \cite{Suzuki}.  The agreement is rather good;  this is particularly impressive in view of the fact that the present model has no adjustable parameters.  (The standard model fits the data even better -- see figure 4 of \cite{Suzuki} -- but has three adjustable parameters: $\Omega_k$, $\Omega_m$, and  $w_{de}$.)  Note that the   fit in  figure \ref{fig1} is not affected by the  Hubble constant   either (since only ratios of distances are measured by the Type Ia supernovae).  That is, the evidence provided by figure \ref{fig1}   has a rare distinction of being ``absolute".

\section{The unexpected behavior of matter density during expansion}\label{density}
The predicted relation between the density of matter and the scale factor, equation (\ref{rho-3/4}), is counterintuitive: it seems obvious that matter density should depend inversely on the volume, ${{\rho }_{m}}\propto {{a}^{-{3}\;}}$. This expectation, however, is based on the non-relativistic law of  conservation of mass, whereas the true conservation law is  that of energy;  this  law -- the first law of thermodynamics -- accounts for the relativistic effects, and leads unambiguously to equation (\ref{rho-3/4}) as shown above.  

Essentially-relativistic results  are usually  counterintuitive, the  famous example being the existence of the rest energy (the mass-energy equivalence). In the present case, the conflict with intuition arises because the cosmic interaction potential  ${{\phi }_{c}}$ is unobservable on the non-cosmological scales  (Section \ref{mean}), yet on the cosmological scales its effect is extremely strong (fully relativistic). Three-fourth of the total energy content of the universe is contributed by the  cosmic interaction, with the stress-energy tensor  ${{\Phi }_{\mu \nu }}$ and pressure that is  equal and opposite to the energy density (equations (\ref{Phi},\ref{rho=4},\ref{p=-3})). For a hypothetical  pure substance  with such an equation of state,  energy density would not change at all with the expansion. From this perspective, it is entirely natural that the energy density of the  cosmic fluid, which includes the  energy density of matter as one-fourth, should change slowly with the expansion --  as equation (\ref{rho-3/4})  indicates. 

The physical meaning of this finding is as follows: As a result  of the cosmic interaction, the expansion of the cosmic fluid, which represents the universe together with its matter content,  cannot be treated as the ``expansion of space'', with  matter passively residing  ``inside''.  On the cosmological level,  space-time and matter are not separable, but form a single entity.

\section{Implications for cosmology II. The growth of structure}\label{growth}
In this section, further comparison of the predictions of the present model with observations will be attempted. It is important to keep in mind  the preliminary nature of this exercise. A definitive comparison will become possible only after a complete cosmology has been built on the basis of the present model. The reason is simple:  interpretation of the raw observational data -- the route from the raw data to the results presented in research literature -- is usually model-dependent. (This  does not apply to the luminosity  data from Type Ia supernovae,  figure \ref{fig1}.)  For example,  analysis of the data often makes use of the  standard relation  ${{\rho }_{m}}\propto {{a}^{-{3}}}$. In the present model, the relation between ${{\rho }_{m}}$ and $a$ is very different -- see equation (\ref{rho-3/4}) and Section \ref{density}. Other model-dependent steps in the data analysis may be more subtle,  but no less consequential. 

With these caveats in mind, the growth of structure in the universe will now be discussed.  Due to gravitational instability, small perturbations in matter density grow with time; the resulting seeds eventually become galaxies. The linear perturbation growth is described by the  equation \cite{Peacock, Mukhanov, Wang} \begin{equation}\label{perturb}\ddot{\delta }+2H\dot{\delta }=4\pi {{\rho }_{m}}\delta ,\end{equation} 
where $\delta (t)$ is the fractional density perturbation. In a static universe ($H = 0$), the growth of perturbations would be exponential; the expansion of the universe slows it down. The growing mode can be sought in the form\begin{equation}\label{alpha}\delta (t)\propto {{t}^{\alpha }};\end{equation} 
substituting this into equation (\ref{perturb}), one finds for the present model\begin{equation}\label{}\alpha =\frac{-13+\sqrt{265}}{6}=0.546.\end{equation} 
By way of comparison, solving equation (\ref{perturb}) for the standard model with $\rho ={{\rho }_{m}}$  (the matter-dominated epoch) yields $\alpha ={2}/{3}\;$.  Note that in both cases\begin{equation}\label{t-2}{{\rho }_{m}}\propto {{t}^{-2}},\end{equation} 
as expected on dimensional grounds.

	Observational data are interpreted in terms of the growth rate $f$, defined by
\begin{equation}\label{f_def}f\equiv \frac{d\log \delta }{d\log a}.\end{equation} 
  
So far, some 20 measurements of $f$  have been made, for redshifts $z<0.8$; the results are compiled in \cite{Tojeiro}. The reported values of $f$  span the range $\sim0.5$ to 0.8, with the majority clustering between 0.6 and 0.7.  The recent analysis  \cite[Fig.~1]{Kwan} suggests that these values are systematically underestimated by about 0.1.  If the corresponding corrections were included, the observational results would cluster between 0.7 and 0.8.

	The present model  cannot be directly compared to the reported measurements because the process of extracting  $f$  from the raw observational data is highly model-dependent.  To disentangle the underlying physics from the model-dependent interpretation is very difficult; normally, a complete reanalysis of the data would be required.  In the present case, however, a simple approximate treatment may suffice, as follows.

Equations (\ref{alpha}, \ref{t-2}) mean that
\begin{equation}\label{indep}\frac{d\log \delta }{d\log {{\rho }_{m}}}=-\frac{\alpha }{2}\end{equation} 
	  
in both cases considered above, i.e., the present model, and the standard model for the matter-dominated epoch.

	The expression (\ref{f_def}) for the growth rate $f$  can be rewritten in the following equivalent form\begin{equation}\label{chain}f=\frac{d\log \delta }{d\log {{\rho }_{m}}}\times \frac{d\log {{\rho }_{m}}}{d\log a}.\end{equation} 
The first factor on the right-hand side is linked directly to the astronomical observables, and its connection with $\alpha$ is expected, on  general grounds, to be independent of the model  (see equations (\ref{t-2}, \ref{indep})). This suggests that  the  effect of using the standard model to interpret the observational data is confined to the second factor. That is, the reported value of $f$  is\begin{equation}\label{}{{f}_{rep}}=\frac{d\log \delta }{d\log {{\rho }_{m}}}\times {{\left( \frac{d\log {{\rho }_{m}}}{d\log a} \right)}_{\Lambda \text{CDM}}}=-3\frac{d\log \delta }{d\log {{\rho }_{m}}}.\end{equation} 
For the present model this yields (see equation (\ref{indep}))\begin{equation}\label{}{{f}_{rep}}=\frac{3\alpha }{2}=0.820.\end{equation} 
Thus the present model is in reasonable agreement with the available measurements of the growth rate.

The Hubble diagram of Type Ia supernovae  (figure \ref{fig1}),  and the growth rate of density perturbations, are the two most important sources of observational evidence for acceleration and dark energy, in other words, for cosmic repulsion. 

\section{Constraints from the cosmic microwave background}
Evidence for  dark energy from the  microwave background is of indirect nature  \cite[\S\S  1.1.3, 2.4.4]{Wang}. The crucial input from the microwave background is the spatial flatness of the universe. The present model is in agreement with this observational result, and provides an explanation for it (Section \ref{curvature}).

	Another important input from the cosmic microwave background is the fraction of matter  in the present-epoch total energy density, ${{\Omega }_{m}}\approx 0.27$ (the rest  attributed to dark energy). The analysis of the microwave background data is based on the standard model \cite{Mukhanov, Amendola}, so this result is potentially model-dependent. However, it is in rough agreement with the estimates of ${{\Omega }_{m}}$ based on several different types of data: luminosities of Type Ia supernovae \cite{Suzuki},  abundance of galaxy clusters, etc. \cite{Wang,Amendola}.  That all these various estimates converge on a single number is an indication that its model-dependence is minimal. The present model is in good agreement with this result as well  --  see equation (\ref{rho=4}).

	In Section \ref{flatacceldark}, the present model was found to be in agreement with all the major conclusions of observational cosmology, which incorporate  constraints from the cosmic microwave background. This is about as far as one can go at the moment. A detailed analysis of the microwave background data will become possible   after a complete cosmology has been built on the basis of the present model.

\section{Global repulsion from pairwise attraction -- a paradox?}
In the present model,  cosmic repulsion arises as a global effect of the attractive gravitational interactions between distant objects. This looks paradoxical: how can global repulsion result from pairwise attraction?  In the van der Waals fluid, attractive interaction also leads to negative pressure (see Appendix), but the overall effect is cohesion, not repulsion. Why does the universe behave in such a counterintuitive way?

	Because (a) in general relativity pressure is a source of gravity, and (b) the cosmic interaction involves the universe as a whole. The negative pressure -- the source of anti-gravity -- attains great magnitude as a result. Global repulsion then dominates the attractive effect of the cosmic interaction (which is also significant -- see equations (\ref{rho},\ref{rho=4})).

	A similar (if sign-reversed) interplay of causes and effects is behind one of the most remarkable predictions of general relativity: the upper bound, $\sim3$ solar masses, on the mass of a neutron star \cite{Rhoades Ruffini}. This arises as follows: As the mass of the star increases, the local values of pressure increase along with the weight of the overlying matter, thus preventing the matter from falling to the center. But positive pressure acts as a source of gravity. Eventually this effect becomes dominant, and the star collapses to a black hole. Note how the short-range repulsion, acting locally to prevent the collapse, leads to the amplified global attraction, and finally to collapse when the latter prevails.

\section{Consistency with the second law of thermodynamics}\label{thermo}
The apparent horizon can be defined as follows: if beyond this horizon a light pulse is emitted toward the origin, its proper distance from the origin increases with time due to the expansion of the universe. In a spatially-flat universe, the radius of the apparent horizon is the Hubble radius ${{H}^{-1}}$; the recession velocity at the apparent horizon equals $c$.  The area of the apparent horizon is ${4\pi }/{{{H}^{2}}}$, and the associated Bekenstein-Hawking entropy is\begin{equation}\label{Sa}{{S}_{a}}=\frac{\pi k}{\hbar {{H}^{2}}},\end{equation} 
where $k$ is the Boltzmann constant, and $\hbar $ is the Planck constant.  (In the units used here, $\hbar $ is equal to the Planck area.)

	The nature and significance  of ${{S}_{a}}$ are not fully understood, but evidence has been accumulating in  favor of its physical reality.  At present, ${{S}_{a}}\sim 2\times {{10}^{122}}k$.  The entropy of the observable universe, $\sim {{10}^{104}}k$, is negligible in comparison \cite{Egan}.  In what follows, I view ${{S}_{a}}$ as the entropy scale of the universe.  Since ${{S}_{a}}$ is fully determined by the instantaneous value $H(t)$ of the Hubble parameter, it is well defined throughout the expansion history.

  	One of the key results of thermodynamics -- the thermodynamic stability condition -- can be stated as follows: The entropy of an isolated system must be a strictly concave function of its arguments (the extensive parameters of the system); the chord joining any two points on the entropy curve must lie below the curve \cite[\S8.1]{Callen}.  This condition is a manifestation of the second law of thermodynamics; its violation entails loss of homogeneity -- the system will increase its entropy by spontaneously separating into subsystems with different properties  \cite[\S8.1]{Callen}.  While local small-scale inhomogeneities (fluctuations) continuously occur in a thermodynamic system, the concavity of the entropy protects the large-scale homogeneity; fluctuations induce processes that resist their growth (the Le Chatelier principle \cite[Ch.~8]{Callen}).

	Observations show that the universe has been homogeneous and isotropic on large scales throughout the expansion history. I conclude that the entropy of the universe is a concave function of its arguments. (Except, possibly, in the radiation-dominated era: radiation does not cluster.)

  	A mathematical statement of concavity ought to involve the characteristic scales of the universe, the entropy scale ${{S}_{a}}$ and the scale factor $a$.  The concavity requirement can thus be stated as follows:\begin{equation}\label{d2Sa}\frac{{{\partial }^{2}}{{S}_{a}}}{\partial {{a}^{2}}}<0\end{equation} 
at all times since the end of the radiation-dominated era.	

	Inequality (\ref{d2Sa}) was first derived by Radicella and Pav\'on  \cite{Radicella},  for the final stage of the expansion only ($a\to \infty $), by  requiring that the universe  tend to a state of maximum entropy.  (They did not consider the homogeneity-concavity connection invoked above.)  Using the Friedmann equations to calculate ${{{\partial }^{2}}{{S}_{a}}}/{\partial {{a}^{2}}}$, they found   that inequality (\ref{d2Sa}) implies the following bounds on the overall equation-of-state parameter $w=p/\rho$ of the cosmological model\begin{equation}\label{-1w-23}-1<w<-\frac{2}{3}.\end{equation} 
According to the view presented here, inequalities (\ref{d2Sa}) and (\ref{-1w-23}) must be satisfied not only during the final stage of the expansion, but at all times since the end of the radiation-dominated era.  Two conclusions follow: (a) the present model, in which  ${-3}/{4}\;\le w<{-2}/{3}$, is consistent with the second law of thermodynamics; (b) the standard model, in which $w \approx 0$  in the matter-dominated epoch, is not.

\section{On the cosmological constant problem}\label{constant}

Neither the cosmological constant nor dark energy enter the present model, so the cosmological constant problem disappears. Or rather transforms into a different problem, as follows.

	Let us recall the essence of the cosmological constant problem. In quantum field theory, the vacuum energy density is estimated to be of the order of the Planck density, i.e.,\begin{equation}\label{vacuum}\text{(vacuum energy density)}\sim {{\hbar }^{-1}}.\end{equation} 
(An alternative possibility  -- not currently favored --  is zero.)  The cosmological constant (dark energy) is thought to be a manifestation of vacuum energy, but its value in the standard model is very much smaller than the estimate (\ref{vacuum}), viz.,\begin{equation}\label{dark}\text{(dark energy density)}\sim {{10}^{-123}}{{\hbar }^{-1}}.\end{equation} 
Thus the problem: Why is the cosmological constant so small?

	In the standard model, the present-epoch total energy density ${{\rho }_{0}}$ is mainly dark energy.  (Why? This is the coincidence problem.)  Hence the last equation can be rewritten as\begin{equation}\label{rho0h}{{\rho }_{0}}\hbar \sim {{10}^{-123}}.\end{equation} 
Consider now the Friedmann equation (\ref{fried1}) in combination with the expression (\ref{Sa}) for the entropy scale of the universe  ${{S}_{a}}$.  The resulting simple relation,\begin{equation}\label{rhohSa}(\rho \hbar )\times ({{{S}_{a}}}/{k}\;)=\frac{3}{8},\end{equation} 
yields immediately the estimate (\ref{rho0h}) for the present epoch (${{S}_{a}}\sim 2\times {{10}^{122}}k$), but is true in general, throughout the expansion history.

	Thus the question ``Why is the cosmological constant so small?'' transforms into the question ``Why is the entropy scale of the universe so large?'' (at present).  An interesting question -- but there is nothing paradoxical about it. 

	At the other extreme, equation (\ref{rhohSa}) implies a very low value for ${{S}_{a}}$ during the Planck epoch, when the energy density of the cosmic fluid would be $\sim {{\hbar }^{-1}}$.

\section{One last paradox vs. one  old problem}\label{oldproblem}

Minkowski (flat) space-time, with ${{g}_{\mu \nu }}=\text{const}\times{{\eta }_{\mu \nu }}$, is a solution of the Einstein field equation (\ref{efe}) in the absence of matter (${{T}_{\mu \nu }}=0$), and a  special  case of a conformally flat space-time.  Thus the considerations of Section \ref{potential}  apply, including the final result for the cosmic potential  ${{\phi }_{c}}$, equation (\ref{-3}).  As long as Minkowski space-time is used to describe (on timescales much shorter than the Hubble time) a finite region of the universe that is devoid of matter, no problems arise -- the cosmic potential  ${{\phi }_{c}}$ in that region  is due to the matter in the rest of the universe.

However, Minkowski space-time is also commonly viewed  as a legitimate  cosmological solution of the Einstein field equation, representing  a hypothetical  empty universe of infinite extent. Such an interpretation now  becomes untenable: In an empty universe, there would be no  source   -- no  cause to exist (Section \ref{mean})  -- for  the cosmic potential  ${{\phi }_{c}}$ required by virtue of equation (\ref{-3}).  A paradox? 

Actually,  the last missing link in  the resolution  of the problem that concerned Newton, Leibniz, and Mach, among others  -- the origin of inertia.  Einstein's view, which guided his work on general relativity \cite[\S2]{Einstein 1916}, is well known: ``inertia [is] to be traced to mutual action with distant masses'' \cite{Einstein 1921}. In general relativity, the geometry of space-time, which determines the local inertia, is itself determined by the distribution of matter in the universe. But there is a  problem:

Minkowski space-time  as a representation of an empty  universe is essentially equivalent to Newton's ``absolute space''.  A body accelerating with respect to the latter feels   resistance -- the force of inertia -- which cannot be attributed to any material cause. Einstein was disturbed by the fact that the field equation of general relativity appeared to admit such a solution, and argued that it must be unphysical \cite[\S2] {Einstein 1917}, \cite{Einstein 1918}. 

 The above ``paradox'' is  a vindication of this position.

\section{Concluding remarks}\label{conclusion}
The model presented in this paper has no adjustable parameters, introduces no additional fields, and eliminates the dark energy concept. The predictions of the model are in good agreement with the major conclusions of observational cosmology. The model resolves three  problems of cosmology -- the flatness problem, the cosmological constant problem, and the coincidence problem. Its consistency with the second law of thermodynamics explains the large-scale homogeneity of the universe.  The cosmic fluid violates the strong energy condition; thus the initial singularity may be avoidable in this model.  Whether the model lives up to its promise remains to be seen.

\section*{Acknowledgments}
\addcontentsline{toc}{section}{Acknowledgments}
I thank Alex Kostinski, Craig Hogan, Andr\'e de Gouv\^ea, Scott Dodelson, Luca Amendola and Donald Lynden-Bell for comments. Figure 1 was kindly provided by  David Rubin.

\section*{Appendix.  Negative pressure in the van der Waals fluid}
\addcontentsline{toc}{section}{Appendix.  Negative pressure in the van der Waals fluid}
Consider a gas of $N$ non-interacting particles in volume $V$ at temperature $T$.  (Here ``non-interacting'' means ``interacting so weakly that equilibrium properties are not affected".)  Its pressure is purely kinetic and obeys the ideal-gas equation of state\begin{equation}\tag{A1}{{p}_{id}}=\frac{NkT}{V}=nkT,\end{equation} 
where  $n = N/V$ is the number density of particles, and $k$ is the Boltzmann constant.  (In this Appendix, $T$ is temperature; in the rest of the paper, $T$ is the trace of the stress-energy tensor.)

When interactions between particles do affect equilibrium properties, they are taken into account as follows. The  repulsive interactions have extremely short range, and are modeled by assuming that the presence of other particles simply reduces the volume available for a test particle, so that\begin{equation}\tag{A2}{{p}_{kin}}=\frac{NkT}{V-Nb},\end{equation} 
 where $Nb$ is comparable to the volume occupied by the same $N$ particles in liquid state.  (The subscript $kin$ stands for ``kinetic''.)  

The range of the attractive interactions between particles is much longer than that of the repulsive ones.  The attractive interactions are taken into account using the mean-field approach. As a preliminary step, consider what happens if the gas is acted upon by some external field with potential $\varphi $.  (Here $\varphi $ is defined per particle and has dimensions of energy.)  If $\varphi $  is non-uniform, each particle is acted upon by a force  $-\nabla\varphi $, and the density and pressure of the gas become non-uniform (as in the barometric formula).  In a uniform external potential -- the only case of interest here -- the free energy of the gas acquires an additive contribution $\varphi N$. 

	In the mean-field approximation, the overall effect of the attractive interaction between a test particle and all other particles is modeled by an effective field with a uniform potential felt by the test particle. The mean-field potential is then expressed, in a self-consistent way, as the integral of the pairwise interaction energy times the number of particles in the volume element, $ndV$. The integration extends from the collision distance to infinity. The integral converges if the pairwise interaction energy falls with the distance faster than ${{r}^{-3}}$; for a typical intermolecular interaction, the dependence is $\sim{{r}^{-6}}$.  With convergence assumed, no further details of the interaction are relevant; the essential result is that the mean-field potential is proportional to the number density of particles. Thus the free energy of the system acquires an additive contribution
\begin{equation}\tag{A3}{{F}_{attr}}= -anN= -a\frac{{{N}^{2}}}{V},\end{equation}  
 where $a>0$.  (In this Appendix, $a$ is the van der Waals constant; in the rest of the paper, $a(t)$ is the scale factor of the universe.) 

	The new contribution to energy is negative because it is potential energy of attractive interaction. If the system expands, this energy increases, approaching zero as $V\to \infty $;  hence the contribution to pressure due to attractive interaction will be negative. Let us denote this contribution, the``attraction pressure'',    by ${{p}_{attr}}$.  (Here  ${{p}_{attr}<0}$;  in physical chemistry, ${-{p}_{attr}}$, a positive quantity, is    called ``internal''  or ``cohesive''  pressure.)  Now  ${{p}_{attr}}$ can be calculated as
\begin{equation}\tag{A4}{{p}_{attr}}= -{\left(\frac{\partial {F}_{attr} }{\partial V}\right)}_{N,T} =-a\frac{{{N}^{2}}}{{{V}^{2}}}=-a{{n}^{2}},\end{equation} 
 
and the total pressure of the gas of interacting particles is\begin{equation}\tag{A5}{{p}_{vdW}}={{p}_{kin}}+{{p}_{attr}}=\frac{NkT}{V-Nb}-a{{\left( \frac{N}{V} \right)}^{2}}.\end{equation} 
Equation (A5) is the van der Waals equation of state. It describes reasonably well the behavior of gases and liquids, and the gas-liquid phase transition. The van der Waals constants $a$ and $b$ depend on the molecular species.

	Dividing the interaction energy  ${-a{{N}^{2}}}/{V}\;$ by the volume of the system, one obtains the mean-field energy density of the attractive interaction, $-a{{n}^{2}}$.  Note that ${{p}_{attr}}$ is equal to this energy density.

	For a gas at room temperature and pressure, ${{p}_{attr}}$  is very small compared to the kinetic pressure ${NkT}/{(V-Nb)}$.  Not so for a liquid under the same conditions. Indeed, if not for ${{p}_{attr}}$, liquids would not exist -- there would be no cohesion.  For example, the volume of one mole of liquid water, 18 $\text{c}{{\text{m}}^{3}}$, is  smaller than the molar volume of  gas by a factor $\sim {{10}^{3}}$.  Thus the kinetic pressure in liquid water is $\sim {{10}^{3}}$ atm. The attraction pressure is equally great (by absolute value) -- the sum of the two equals 1 atm. 

	When liquids are put under tension, the attraction pressure exceeds the kinetic pressure (by absolute value). The liquid is then in a metastable state, and considerable efforts are required to prevent the formation of vapor bubbles; nevertheless, tensions $\sim 300$ atm have been achieved in experiments.

\end{document}